\documentclass[11pt]{article}
\usepackage{graphicx}
\DeclareGraphicsExtensions{.eps,.ps}
\usepackage{epsfig, graphicx}

\usepackage{subfigure}
\newcommand{\be}[1]{\begin{equation} \label{eq#1} }
\newcommand{\ee}{\end{equation}}
\newcommand{\bea}[1]{\begin{eqnarray} \label{eq#1} }
\newcommand{\eea}{\end{eqnarray}}

\def\setb@se#1{\baselineskip=#1 \normalbaselineskip=#1}
\def\captionfont{\setb@se{10pt}\protect\footnotesize}
\begin{document}
\mbox{}

\vskip0.0cm
\begin{center}

{\large Analysis on the Coupling Effects between Elastic and Electromagnetic Fields from the Perspective of Conservation of Energy \\}
\vskip0.7cm  Peng Zhou$^1$\\[0.4cm]
$^1$Department of Mechanics, Harbin Institute of Technology, Harbin, \\
    Heilongjiang, 150001, P.R. China  \\
    Email: zhoup@hit.edu.cn
\vskip0.0cm

\end{center}
\vskip0.4cm

\begin{abstract}
Coupling effects among different physical fields substantially reflect the conversion of energies from one form into another. For simple physical processes, their governing or constitutive equations all satisfy the law of conservation of energy. Then, analysis is extended to coupling effects. First, it is found for the linear direct and converse piezoelectric and piezomagnetic effects, their constitutive equations guarantee that the total energy is conserved during the process of energy conversion between the elastic and electromagnetic fields; however, energies are converted via work terms, $(\beta_{ijk} E_i )_{,k} v_j$ and $(\gamma_{ijk} H_i)_{,k} v_j$, rather than via energy terms, $\beta_{ijk} E_i e_{jk}$ and $\gamma_{ijk} H_i e_{jk}$. Second, for the generalized Villari effects, the electromagnetic energy can be treated as an extra contribution to the generalized elastic energy. Third, for electrostriction and magnetostriction, it is argued both effects are induced by the Maxwell stress; moreover, their energy is purely electromagnetic and thus both have no converse effects. During these processes, energy can be converted in three different ways, i.e., via non-potential forces, via cross-dependence of energy terms and directly via the electromagnetic interactions of ions and electrons. In the end, general coupling processes which involve elastic, electromagnetic fields and diffusion are also analyzed. The advantage of using this energy formulation is that it facilitates discussions of the conversion of energies and provides better physical insights into the mechanisms of these coupling effects.
\end{abstract}

\vspace{0.7cm}


\newpage

\vspace{10cm}

Revisions: More physical explanations, indicated by \textbf{bold fonts}, are provided for the process of energy conversion during the linear piezoelectric and piezomagnetic effects in section 3. This portion of materials has been presented during the 14th USNNCM at Montreal, Canada (07/2017). A schematic illustration of the conversion of energies for the three coupling effects between electromagnetic fields and elastic fields are also summarized  in Figure 2 at the end of section 5.

\newpage

\section {Introduction}

The study of coupling effects between electromagnetic and elastic fields dates back to more than one century.$^{\cite{Tichy:2010, Sundar:1992}}$ The first set of such coupling effects are piezoelectric and piezomagnetic effects. Piezoelectricity was first reported by the Curie brothers in literatures at 1881.$^{\cite{Curie:1881}}$ The direct piezoelectric effect refers to a change of polarization in certain substances due to externally applied stresses; while the converse effect refers to the arising of strains due to externally applied electric fields. These effects only exist in materials with microstructures showing no inversion of symmetry. Piezoelectric effects were initially discovered in crystals such as tourmaline and quartz, and latter in ceramics. Nowadays, they are also found in polymers and even biological materials such as tendons and bones. Due to persistent efforts in searching for better materials and in understanding the effects both microscopically and macroscopically, it eventually led a wide application of these effects in industrial manufacturing, medical instruments and telecommunication devices. Piezomagnetism is the magnetomechanical analogue of piezoelectricity, i.e., externally applied stresses results in a change of magnetization and externally applied magnetic fields lead to arising of strains. No inversion of symmetry is also required for the microstructure of piezomagnetic materials. In 1960, this effect was experimentally observed in antiferromagnetic fluorides of cobalt and manganese.$^{\cite{Romanov:1960}}$ The second set of these coupling effects are electrostriction and magnetostriction. Both of them refer to dimensional changes of materials under the influence of externally applied electric or magnetic field. Magnetostriction was reported by Joule in iron in 1842.$^{\cite{Joule:1842}}$ There are two specific effects of magnetostriction called the Matteucci effect$^{\cite{Matteucci:1858}}$ and the Wiedemann effect.$^{\cite{Wiedemann:1862}}$ The former refers to the arising of a helical anisotropy of the susceptibility in magnetic materials when subjecting to a torque; while the latter the arising of a torque within the bulk of materials when subjecting to a helical magnetic field. In addition, there is another effect called the Villari effect, or the inverse magnetostrictive effect. It refers to the phenomenon that an externally applied stress induces a change in the magnetic susceptibility of a magnetic material.$^{\cite{Villari:1865}}$  Moreover, it is worth noting that strains induced for the first set are linearly dependent on externally applied electromagnetic fields; while those for the second set are quadratic. As a result, a change of the field directions leads to a change of the sign of strains for the former but not the latter.

As shown above, the coupling of the electromagnetic and elastic fields leads to a variety of effects. The underlying mechanisms of these effects are both important and interesting. Many models and theoretical approaches were developed over the past century to study them both microscopically and macroscopically. To be very brief here, the microscopic understanding mainly focuses on the relations between the crystal structures and properties of the materials, while the macroscopic understanding mainly uses phenomenological models to obtain the constitutive equations of these coupling effects. One major theoretical approach at the macroscopic level is the Lagrangian formulation. The Lagrangian formulation was initially developed in analytical mechanics and latter extended to continuum mechanics to derive the equations of motion. In electrodynamics, Lagrangian formulation can also be used to derive governing equations for the scalar and vector potentials for the electric and magnetic fields. These equations are in fact equivalent to the set of Maxwell equations. Using the Lagrangian formulation, governing equations are derived to ensure that the first variation of the integral of Lagrangian, which is defined to be the difference between the kinetic and potential energies of the system, over a time interval is zero. A detailed theory of piezoelectricity and its application in piezoelectric devices can be found in literatures such as here.$^{\cite{Tichy:2010}}$ However, given the fact that the coupling effects between electromagnetic and elastic fields essentially arises from the energy conversion between the electromagnetic fields and the elastic field, it appears that one more important issue is still left unaddressed. That is, how is energy converted between these two fields and how can this process be formulated?

In this paper, the fundamental law of conservation of energy is applied to analyze these coupling effects. For brevity, this approach is hereafter called the energy formulation. It is shown this energy formulation enables the governing equations of these coupling effects to strictly satisfy the law of conservation of energy. The purpose of this paper is two-fold: first, using the law of conservation of energy to analyze these coupling effects and thus, to contribute to a straightforward understanding of the conversion of energies during these effects and to provide more physical insights into their mechanisms; second, using these examples to show that the energy formulation can be developed into a general approach to analyze the coupling effects among reversible and irreversible processes. Organization of this paper is as follows. In the second section, a general discussion on the relations between the governing equations and the law of conservation of energy for simple processes is provided. In the third and fourth sections, the piezoelectric and piezomagnetic effects, as well as the Villari effect and its analogue in the electric field are analyzed, respectively. In the fifth section, electrostriction and magnetostriction are studied. In the sixth section, general coupling processes involving elastic, electromagnetic fields and diffusion are analyzed. In the seventh section, some comments on the energy conversion is provided and a summary is given in the end.

\section{Governing Equations and the Law of Conservation of Energy}

In this section, the relation between the governing equations and the law of conservation of energy for uncoupled and simple processes is discussed. It is shown not only in classic mechanics but also in thermodynamics and electrodynamics, the governing or constitutive equations must satisfy the law of conservation of energy.

For simple processes, to begin with, the equation of motion in classic mechanics is discussed. In a natural system with the kinetic energy $T(\dot{q})$ being a homogenous quadratic function of the generalized velocity $\dot{\mathbf{q}}$, the Lagrangian is defined to be $ L = T(\dot{\mathbf{q}})-V(\mathbf{q})=\frac{1}{2} m_{ij} \dot{q_i} \dot{q_j}-V(q_k)$, where $V(q_k)$ is the potential energy and the generalized mass $m_{ij}$ is a symmetric tensor. According  to analytical mechanics,$^{\cite{Leonard:1970}}$ in this system, Lagrange's equations or the equations of motion, $\frac{d}{dt}(\frac{\partial L} {\partial \dot{q_k}}-\frac{\partial L} {\partial q_k}) = 0$, lead to the law of conservation of energy directly. That is, $
  [\frac{d}{dt}(\frac{\partial L} {\partial \dot{q_k}})-\frac{\partial L} {\partial q_k}]\dot{q_k} = \frac{d}{dt}(T+V) = 0$. This in fact provide an alternative approach to determine the equations of motion for a mechanical system. When the total energy $E=T+V$ of this system is determined, the time derivative of $E$ can be written as a linear combination of the generalized velocities $\dot{q_k}$. Then the coefficients associated with each $\dot{q_k}$ must be zero, which give the equations of motion, because the series $\dot{q_k}$ are linearly independent.

In analytical mechanics, the Lagrangian formulation is a standard approach to determine the equations of motion in a mechanical system. But there is one major disadvantage with this formulation. The equations of motion are derived to minimize a scalar integral of the Lagrangian over a time interval. Thus, they do not offer any physical insights into the issue of conversion of energies. In the following discussions, it is straightforward to show that the energy formulation can conveniently achieve this purpose. The time derivative of $E$ gives
\begin{eqnarray}
 \frac{dE}{dt} =\frac{d(T+V)}{dt}= [m_{ij}\ddot{q_i} - (- \frac{\partial V}{\partial q_j})] \dot{q_j}
\end{eqnarray}
where the equations of motion are found by setting the coefficients in the square bracket to be zero. Note that, the first term in the square bracket, $m_{ij}\ddot{q_i}$, is the inertia force; a positive product of it with the generalized velocity $\dot{q_j}$ indicates an increase of the kinetic energy $T$, which gives the rate of conversion of the potential energy $V$ into $T$. In the second term, $- \frac{\partial V}{\partial q_j}$ is the potential force; a positive product of it with $\dot{q_j}$ indicates a positive work and a decrease in the potential energy $V$, which also gives the rate of conversion of the potential energy $V$ into $T$. Thus, these two terms must cancel each other, which leads to the law of conservation of energy. It is evident that, with the energy formulation, the conversion of energies can be discussed more straightforwardly.

The energy formulation can also be extended to continuum mechanics to give the Cauchy's equation of motion. In a continuum media, the total energy can be defined as $\int_V [\frac{1}{2} \rho v_i v_i + u^{el}(e_{ij})] dv$ where $\rho$ is the mass density, $v_i$ is the velocity and $u^{el}(e_{ij})$ is the elastic energy density function of the infinitesimal elastic strain $e_{ij}$. Then,
\begin{eqnarray}
 \frac{dE}{dt} &=& \frac{d}{dt} \int_V [\frac{1}{2} \rho v_i v_i + u^{el}(e_{ij})] dv \nonumber \\
       &=& \int_V [\frac{\partial}{\partial t}(\frac{1}{2} \rho v_i v_i) + \frac{\partial u^{el}}{\partial e_{ij}} \frac{\partial e_{ij}}{ \partial t} ] dv \nonumber \\
       &=& \int_V (\rho v_i \frac{\partial v_i}{\partial t}  + \sigma_{ij} v_{i,j} ) dv \nonumber \\
       &=& \oint_A \sigma_{ij} n_j v_i da + \int_V (\rho \frac{\partial v_i}{\partial t} - \sigma_{ij,j} ) v_{i} dv,
\end{eqnarray}
where at the second step, the elastic stress $\sigma_{ij}=\frac{\partial u^{el}}{\partial e_{ij}}$, $\frac{\partial e_{ij}}{ \partial t}=\frac{1}{2}(v_{i,j}+v_{j,i})$ and the symmetry of indices $i$ and $j$ is used. Thus, to guarantee the energy is conserved in a continuum media with a stress-free boundary, the terms in the parenthesis of the volume integral at the last step must be zero, which leads to Cauchy's equation of motion directly. Here, the advantage of the energy formulation is also evident when it comes to the discussion of conversion of energies. In the last step, the surface integral indicates the work done on the continuum media via surface traction $t_i = \sigma_{ij} n_j$ and the surface density of work is $t_i v_i$. Within the volume integral, the positive products of both the inertia force $\rho \frac{\partial v_i}{\partial t}$ and the potential force $\sigma_{ij,j}$ with the velocity $v_i$ indicate the rate of conversion of the elastic potential energy $u^{el}$ into the kinetic energy, and thus they are canceled.

In the above discussions, it is shown that in both classic and continuum mechanics, the law of conservation of energy can be used to determine the equation of motion. Compared with the Lagrangian formulation, the energy formulation directly facilitate the discussion of conversion of energies. In the following discussions, the energy formulation is extended to study the governing equations in both thermodynamics and electrodynamics.

The governing equations in thermodynamics are diffusion-type equations. Recently, the phase-field variational approach (PFVA) has become a popular tool to construct governing equations which abide by the corollary of the second law of thermodynamics, i.e., the total free energy is non-increasing during the evolution of an isolated system.$^{\cite{Penrose:1990, Wang:1993}}$ A brief outline of this approach is presented here. Suppose in a thermodynamic system, the chemical free energy density is $f(c)$ and the chemical potential is $\mu=\frac{\partial f}{\partial c}$. Then,
\begin{eqnarray}
 \frac{d}{dt} \int_V f(c) dv &=& \int_V \frac{\partial f}{\partial c} \frac{\partial c}{ \partial t}  dv   \\
       &=&  - \int_V \mu \partial_i J^c_i  dv \nonumber \\
       &=& - \oint_A \mu  J^c_i n_i da + \int_V \partial_i \mu  J^c_{i} dv, \nonumber
\end{eqnarray}
where the law of conservation of mass $\frac{\partial c}{ \partial t}=-\partial_i J^c_i$ is applied in the second step. Then to guarantee in an isolated system the total free energy is non-increasing as time evolves, i.e., the satisfaction of the corollary of the second law of thermodynamics, the compositional flux $J^c_i$ needs be defined as $J^c_i = - m \partial_i \mu$, where $m$ is the mobility of diffusion. Considering that during the thermodynamic evolution, the chemical free energy is usually dissipated as heat, then a slight modification by introducing a dissipation function in the above derivation can guarantee the first law of thermodynamics is also satisfied. Let the rate of change of the free energy dissipation function be $\frac{1}{m} J^c_i J^c_i$, which is in analogue with that considered in classic mechanics,$^{\cite{Leonard:1970}}$ then the variation of the total energy is
\begin{eqnarray}
 \frac{dE}{dt}  &=& \int_V \frac{1}{m} J^c_i J^c_i dv + \frac{d}{dt} \int_V f(c) dv \\
       &=& - \oint_A \mu  J^c_i n_i da + \int_V (\frac{1}{m} J^c_i +\partial_i \mu)  J^c_{i} dv. \nonumber
\end{eqnarray}
Thus in an isolated system, to guarantee the satisfaction of the law of conservation of energy, the constitutive flux equation is found to be $J^c_i = - m \partial_i \mu$. As a result, the energy formulation combined with PFVA can be used to determine the equation of flux, which satisfy the law of conservation of energy, for a simple and uncoupled thermodynamic process.

In electrodynamics, the governing equations are the set of Maxwell equations:
\begin{eqnarray}
 \partial_i D_i &=& \rho, \label{eq:me1}\\
 \xi_{ijk} \partial_j E_k &=& - \partial_t B_i, \label{eq:me2}\\
 \partial_i B_i &=& 0, \label{eq:me3}\\
 \xi_{ijk} \partial_j H_k &=& J^e_i + \partial_t D_i, \label{eq:me4}
\end{eqnarray}
where $D_i$ ($B_i$) is the electric displacement (magnetic induction); $E_i$ ($H_i$) is the electric (magnetic) field; $\xi_{ijk}$ is the permutation symbol; $J^e_i$ is the free electric current density. Among them, Eqns (\ref{eq:me1}) and (\ref{eq:me3}) are time-independent, i.e., static equations; while Eqns (\ref{eq:me2}) and (\ref{eq:me4}) are time-dependent, i.e., dynamic equations.  Here, $D_i=\epsilon_{ij} E_j$ ($B_i=\mu_{ij} H_j$) where $\epsilon_{ij}$ ($\mu_{ij}$) is the permittivity (permeability). The law of conservation of energy in electrodynamics is called the Poynting's theorem. A brief outline$^{\cite{Jackson:1999}}$ is presented below. Beginning with the work done by the electric field $J^e_i E_i$,
\begin{eqnarray}
 J^e_i E_i  &=& (\xi_{ijk} \partial_j H_k - \partial_t D_i) E_i + (-\xi_{ijk} \partial_j E_k - \partial_t B_i) H_i  \label{eq:em} \\
       &=& -\partial_i(\xi_{ijk} E_j H_k) - \epsilon_{ij} E_i \partial_t  E_j -  \mu_{ij} H_i \partial_t H_j \nonumber  \\
       &=& -\partial_i(\xi_{ijk} E_j H_k) - \partial_t (\frac{1}{2} \epsilon_{ij} E_i E_j + \frac{1}{2} \mu_{ij} H_i H_j ) \nonumber  \\
       &=& -\partial_i S_i - \partial_t u^{em} \nonumber \\
\Longrightarrow -\partial_i S_i &=& \partial_t u^{em} + J^e_i E_i \label{eq:emp}
\end{eqnarray}
 where $S_i=\xi_{ijk} E_j H_k$ is the Poynting vector and $u^{em} = \frac{1}{2} \epsilon_{ij} E_i E_j + \frac{1}{2} \mu_{ij} H_i H_j $ is the electromagnetic energy. At the first step, both the dynamics equations are substituted. Here, on the right-hand side of Eqn (\ref{eq:emp}), the first term indicates the work done by the electric field; the second term, analogously, can be considered as the work done by the magnetic field. Since in the expression of the Lorentz force, the magnetic force in fact does no work, then this term is zero. Note, here the magnetic force is the potential force resulting from the vector potential of the magnetic field. The nonpotential damping force during the magnetization process is not considered here. Note that at the third step and hereafter, the symmetry of the permittivity tensor $\epsilon_{ij}$ and the permeability tensor $\mu_{ij}$ are used so that $u^{em}$ can be explicitly defined. At the last step of Eqn (\ref{eq:emp}), the physical content of Poynting's theorem states the energy transferred into the boundary of a continuum media by the Poynting vector (usually from the electric power source) equals the sum of the increase in the electromagnetic energy and the work done by the electric field (which is usually dissipated into heat or converted into other energies via electronic devices) within the bulk of the material. Shown by the above derivation, it is evident that the two dynamic equations of Maxwell equations must satisfy the law of conservation of energy.

The Lagrangian formulation has already been extended to study electrodynamics and the Maxwell equations can be derived from it. However, the Lagrangian formulation also does not provide any physical insight into the conversion of energies here. Furthermore, the form of the Lagrangian is also a bit puzzling since the electrical and magnetic potentials are assigned with opposite signs. Compared with the law of conservation of energy in the above two cases, Poynting's theorem is of a very different form. However, if starting with the time derivative of $u^{em}$ and the divergence of the Poynting vector as shown in the third step of Eqn (\ref{eq:em}), then the two dynamic equations can still be found by identifying the electric and magnetic work as shown in the first step of Eqn (\ref{eq:em}). That is, in electrodynamics, the energy formulation can also be used to determine the two dynamic equations of the Maxwell equations via Poynting's theorem.

In the above discussions, it is argued that the governing equations of uncoupled and simple processes in mechanics, thermodynamics and electrodynamics can be determined using the law of conservation of energy, i.e., the energy formulations. However, most natural processes are coupled and complex processes, which usually involves two or even more physical fields. During these coupling processes, variation of one field usually leads to variation of another field simultaneously. Thus, it results in conversion of energies from one field into another. In our previous work, it was shown that for coupled irreversible thermodynamic processes,  the energy is in fact conserved during the process of conversion of energies using the flux equations found via PFVA with the free energy functionals. In the following sections, using concrete examples such as piezoelectric and piezomagnetic effects, further discussions are provided for coupled reversible processes between elastic and electromagnetic fields. It is shown that governing equations found with the energy formulation strictly satisfy the law of conservation of energy. Thus, discussion of conversion of energies during these coupling processes becomes straightforward.

\section{Linear Piezoelectric and Piezomagnetic Effects}

For the direct piezoelectric and piezomagnetic effects which behave linearly, their constitutive equations are
\begin{eqnarray}
D_i = \epsilon_{ij} E_j + \beta_{ijk} e_{jk}  \label{eq:di} \\
B_i = \mu_{ij} H_j + \gamma_{ijk} e_{jk}  \label{eq:bi}
\end{eqnarray}
where $\beta_{ijk}$ ($\gamma_{ijk}$) is the piezoelectric (piezomagnetic) coefficient and $e_{jk}$ is the elastic infinitesimal strain tensor. For the converse effect, the constitutive equation is
\begin{eqnarray}
\sigma^{ge}_{jk} = C_{jklm} e_{lm} +  \beta_{ijk} E_i +  \gamma_{ijk} H_i, \label{eq:pe}
\end{eqnarray}
where $\sigma^{ge}_{jk}$ is the generalized stress tensor, $C_{jklm}$ is the stiffness tensor and $\sigma^{el}_{jk} = C_{jklm} e_{lm}$ is the elastic stress tensor.

\textbf{In constitutive equations, the elastic stresses $\sigma^{el}_{jk}$ can also be chosen as independent variables. Then using $e_{jk} = S_{jklm} \sigma^{el}_{lm}$, the above constitutive equations become}
\begin{eqnarray}
D_i &=& \epsilon_{ij} E_j + \beta_{ijk} S_{jklm} \sigma^{el}_{lm} = \epsilon_{ij} E_j + \beta'_{ilm} \sigma^{el}_{lm} \nonumber \\
B_i &=& \mu_{ij} H_j + \gamma_{ijk} S_{jklm} \sigma^{el}_{lm} =  \mu_{ij} H_j  + \gamma'_{ilm} \sigma^{el}_{lm}  \nonumber \\
e^{ge}_{st} &=& S_{stjk} \sigma^{ge}_{jk} = S_{stjk} \sigma^{el}_{jk} +  S_{stjk} \beta_{ijk} E_i +   S_{stjk} \gamma_{ijk} H_i  \nonumber \\
&&\Longrightarrow e^{ge}_{st} = S_{stjk} \sigma^{el}_{jk} +  \beta'_{ist} E_i +  \gamma'_{ist} H_i \nonumber \\
&& \hspace{0.7cm} e^{ge}_{jk} = S_{jklm} \sigma^{el}_{lm} +  \beta'_{ijk} E_i +  \gamma'_{ijk} H_i \nonumber
\end{eqnarray}
\textbf{Note that $e^{ge}_{jk}$ is the generalized strain tensor; $S_{jklm}$ is the compliance tensor and the symmetry of its indices is used in the above derivations. }

\textbf{However, using elastic strains, i.e., $e_{jk}$, as independent variables helps to yield straightforward physical interpretations for each term in the derivations below. Thus, in the following analyzes, the constitutive equations (\ref{eq:di}), (\ref{eq:bi}), and (\ref{eq:pe}) are used; furthermore, it is shown that, beginning with the constitutive equations for the direct effects, the law of conservation of energy leads to the constitutive equations for the converse effects directly; finally and most importantly, discussion on the conversion of energies becomes straightforward.}

Beginning with the work done by the electric field,
\begin{eqnarray}
J^e_i E_i &=& (\xi_{ijk} \partial_j H_k - \partial_t D_i) E_i + (-\xi_{ijk} \partial_j E_k - \partial_t B_i) H_i  \nonumber \\
&=& -\partial_i (\xi_{ijk} E_j H_k ) - E_i \epsilon_{ij} \partial_t E_j-H_i \mu_{ij} \partial_t H_j - E_i \beta_{ijk} \partial_t e_{jk} - H_i  \gamma_{ijk} \partial_t e_{jk} \nonumber \\
&=& -\partial_i (\xi_{ijk} E_j H_k ) - \partial_t [ \frac{1}{2}( \epsilon_{ij} E_i  E_j+ \mu_{ij} H_i H_j) ]  - E_i \beta_{ijk} \partial_t e_{jk} - H_i  \gamma_{ijk} \partial_t e_{jk}, \\
\Longrightarrow  -\partial_i S_i & = & \partial_t u^{em} + J^e_i E_i + E_i \beta_{ijk} \partial_t e_{jk} + H_i  \gamma_{ijk} \partial_t e_{jk} \label{eq:pz}
\end{eqnarray}
where at the second step, the constitutive equations of the direct piezoelectric and peizomagnetic effects are substituted. At the last step, it is straightforward to see that the last two terms representing the conversion of energies due to the coupling effects. Considering the converse effects during which the energy, transferred by the Poynting's vector from the power source, is converted into the elastic energy $u^{el}$. Then the rate of the increase in $u^{el}$ is $E_i \beta_{ijk} \partial_t e_{jk} + H_i  \gamma_{ijk} \partial_t e_{jk}$ according to Eqn (\ref{eq:pz}). Thus, in this case, the conservation of the total mechanic energy is
\begin{eqnarray}
 \frac{d K }{dt} + \frac{d U^{el}}{dt} + \int_V ( E_i \beta_{ijk} \partial_t e_{jk} + H_i  \gamma_{ijk} \partial_t e_{jk} ) dv = 0,
\end{eqnarray}
where the total kinetic energy $K = \int_V \frac{1}{2} \rho v_j v_j dv$ and the total elastic energy $U^{el} = \int_V u^{el} dv= \int_V \frac{1}{2}C_{lmst}e_{lm}e_{st} dv$. Then,
\begin{eqnarray}
 && \int_V [ \partial_t(\frac{1}{2} \rho v_j v_j) + \frac{\partial u^{el}}{\partial e_{jk}} \partial_t e_{jk} +  E_i \beta_{ijk} \partial_t e_{jk} + H_i  \gamma_{ijk} \partial_t e_{jk} ] dv  \nonumber  \\
&=& \int_V [ \rho v_j \partial_t v_j + (\frac{\partial u^{el}}{\partial e_{jk}}  +  \beta_{ijk} E_i +  \gamma_{ijk} H_i) \partial_t e_{jk} ] dv \nonumber  \\
&=& \int_V [ \rho \partial_t v_j - (\sigma^{el}_{jk}  +  \beta_{ijk} E_i +  \gamma_{ijk} H_i)_{,k} ] v_j dv + \oint_A (\sigma^{el}_{jk}  +  \beta_{ijk} E_i +  \gamma_{ijk} H_i) v_j n_k da  \\
&=& 0
\end{eqnarray}
Thus, the equation of motion, taking into account the coupling effects, is
\begin{eqnarray}
\rho \partial_t v_j - (\sigma^{el}_{jk}  +  \beta_{ijk} E_i +  \gamma_{ijk} H_i)_{,k} = 0,
\end{eqnarray}
and the generalized stress is found to be
\begin{eqnarray}
\sigma^{ge}_{jk} = \sigma^{el}_{jk}  +  \beta_{ijk} E_i +  \gamma_{ijk} H_i,
\end{eqnarray}
which is exactly the constitutive equation (\ref{eq:pe}) for the converse effects. That is, the constitutive equations for the direct and converse effects guarantees the satisfaction of the law of conservation of energy.

The terms indicating the conversion of energies between the elastic and electromagnetic fields are
\begin{eqnarray}
  && \int_V [ (\beta_{ijk} E_i +  \gamma_{ijk} H_i) \partial_t e_{jk} ] dv  \\
&=& \int_V [ - (\beta_{ijk} E_i +  \gamma_{ijk} H_i)_{,k} ] v_j dv + \oint_A ( \beta_{ijk} E_i +  \gamma_{ijk} H_i) v_j n_k da, \label{eq:np}
\end{eqnarray}
which consists of a bulk contribution and a surface contribution. Assuming a homogeneous distribution of the electromagnetic fields within the bulk of the devices, the surface contribution becomes dominant. Then, the above equation can be used to obtain the rate of conversion of energies for a certain device with a specific shape, directly. 

\textbf{But note that in Eqn (\ref{eq:pz}), due to the lack of terms, $\beta_{ijk}  e_{jk} \partial_t E_i + \gamma_{ijk} e_{jk}\partial_t H_i$, the variation of energies owing to the piezoelectric and piezomagnetic effects can not be written as the time derivatives of energy terms such as $\beta_{ijk} E_i e_{jk}$ and $\gamma_{ijk} H_i e_{jk}$. On contrast, the elastic and electromagnetic energies are explicitly shown in terms such as $\partial_t u^{el}$ and $\partial_t u^{em}$ in the above derivations. Moreover, when an energy term is a function of vectors or tensors, it is usually a positive-definite quadratic function of its independent variables. The elastic and electromagnetic energies are such examples. Hence, whenever the strains switch signs or the electric and magnetic fields switch directions, these energies remain positive. However, terms such as $\beta_{ijk} E_i e_{jk}$ and $\gamma_{ijk} H_i e_{jk}$ change signs when the strains switch signs or the electric and magnetic fields switch directions. These terms are odd functions of $E_i$, $H_i$ and $e_{jk}$; thus, they introduce a dependence of energy terms on the directions of physical fields, which is also not reasonable. In addition, the elastic energy comes from the work done by externally applied forces; the electromagnetic energy comes from the energy transferred by the Poynting's vector from the power source. However, in the traditional approach, there is no clue about the origin of these piezoelectric and piezomagnetic ``energy" terms. More importantly, the phenomenological assumption of these ``energy" terms also does not help to clarify the following critical problem. That is, how is energy converted between the elastic and electromagnetic fields? Finally, in the traditional approach, the total free energy for the piezoelectric or piezomagnetic materials usually consist of a positive elastic energy term, and a negative electromagnetic energy term with a negative piezoelectric or piezomagnetic ``energy" term. Assigning these terms with opposite signs is also not justified, since the total free energy should be a summation of all three terms. }

\textbf{Thus, it is argued here, for piezoelectric and piezomagnetic effects, there are no specific energies associated with these two processes. Note that, in classic and fluid mechanics, not all processes are associated with specific energies, e.g., the processes due to friction and viscosity which dissipate the kinetic energy of mechanic movements into heat. These forces, usually called nonpotential forces (i.e., they can not be derived from potential functions), are associated with work which convert energies from one form into another. As a result, it is argued here, the variation of energies owing to the piezoelectric and piezomagnetic effects are in fact work terms, $(\beta_{ijk} E_i +  \gamma_{ijk} H_i)_{,k} v_j$, as shown in Eqn (\ref{eq:np}), with their nonpotential body forces being $(\beta_{ijk} E_i)_{,k}$ and $(\gamma_{ijk} H_i)_{,k}$. Note that, these forces are not proportional to velocities as in the case of the viscous force, because they are responsible for conversion of energies rather than dissipation of energies into heat. In brief, the piezoelectric and piezomagnetic effects in fact exemplify the conversion of energies owing to the work done by nonpotential forces.}

\begin{figure}[htb]
\centering 
\epsfig{figure=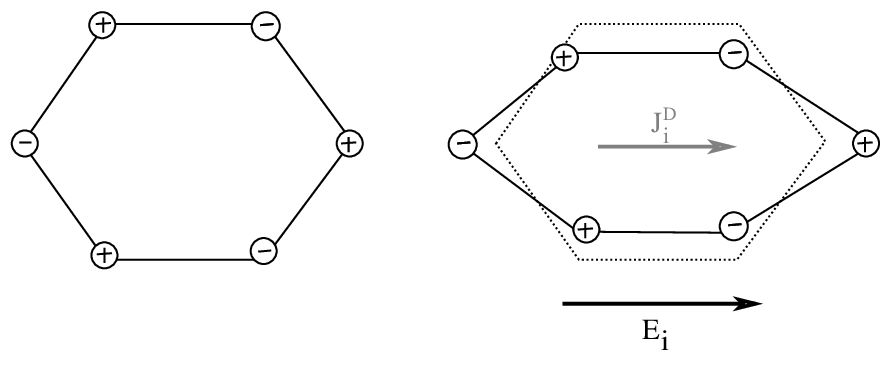,width=0.8\textwidth}
\caption{A schematic diagram of the work done on the displacement currents $J^D_i$ by $E_i$ results in a distortion of the crystal lattice during the converse piezoelectric effect.}
\label{fig:pzee}
\end{figure}

\textbf{Then how to understand the conversion of energies via nonpotential forces intuitively, careful examinations of the physical contents of these work terms are still needed. First, Eqn (\ref{eq:pz}) can be rewritten as }
\begin{eqnarray}
 -\partial_i S_i & = & E_i \epsilon_{ij} \partial_t E_j+ H_i \mu_{ij} \partial_t H_j  + J^e_i E_i + E_i \beta_{ijk} \partial_t e_{jk} + H_i  \gamma_{ijk} \partial_t e_{jk} \nonumber \\
 &=& E_iJ^e_i+ E_i \partial_t(\epsilon_{ij}E_i + \beta_{ijk}e_{jk}) + H_i \partial_t (\mu_{ij} H_j + \gamma_{ijk} e_{jk}) \nonumber \\
 &=& = E_i J^e_i + E_i \partial_t D_i + H_i \partial_t B_i .
\end{eqnarray}
\textbf{At the last step of the above equation, it can be seen the energy transferred from the power source by the Poynting vector is used to do work by both the electric field and the magnetic field. For the electric field, the first term, $E_i J^e_i$, is the work done on the free electric currents. This portion of work is usually converted into Joule heating, or used to achieve electromigration and etc.. The second term, $E_i \partial_t D_i$, is the work done on the displacement currents. Here $\partial_t D_i=J^D_i$ is the displacement current density. The third term, $H_i \partial_t B_i$, is the work done by the magnetic field.  At the absence of piezoelectric and piezomagnetic effects, the second and third work terms are converted or stored as the electromagnetic energy, $u^{em}$. However, when the above two effects are present, extra work are done by both the electric and magnetic field as shown by $E_i \partial_t(\beta_{ijk}e_{jk}) + H_i \partial_t (\gamma_{ijk} e_{jk})$. These are in fact  the integrand in the left side of Eqn (\ref{eq:np}), thus they are both work terms. Here use the converse piezoelectric effect as an example.  An externally applied electric field leads to polarization in the peizoelectric materials and polarization results in displacement currents. Then, work is done on these displacement currents by the electric field. The term $E_i \partial_t(\beta_{ijk}e_{jk})$ is actually a port of  this work, i.e., $E_i J^D_i$, done on the displacement currents by the electric field. During polarization of the peizoelectric materials, besides the electronic polarization, ionic polarization also arises from the deviation of these ions away from their equilibrium positions. This usually causes distortion of the crystal lattice as shown schematically in Figure ~\ref{fig:pzee}. Thus, strains and stresses arise which lead to an increase in the elastic energy within the bulk of the materials. As a results, a port of the work done on  the displacement currents by the electric field is converted into the elastic energy of the materials. This is the underlying mechanism of the peizoelectric effects and they arise from the above mentioned work rather than any specific energy. Here, both the work done by $E_i$ on the free currents $J^e_i$ and on the portion of displacement currents $\partial_t (\beta_{ijk}e_{jk})$ can not be related to or derived from any specific energies. They are just work done by nonpotential forces and they help to achieve conversion of energies between different physical fields.}

\section{The generalized Villari Effects}

In this section, the Villari effect and its analogue in electric field are analyzed. It is reasonable to assume that there is an analogue in the electric field, i.e., the permittivity tensors also vary under the influence of an externally applied stresses. These effects are hereafter called the generalized Villari effects. To facilitate the discussion, here the permittivity and permeability tensors are treated as functions of strains, i.e., $\epsilon_{ij}(e_{kl})$ and $\mu_{ij}(e_{kl})$. Then according to Poynting's Theorem,
\begin{eqnarray}
J^e_i E_i &=& -\partial_i S_i - \partial_t [ \frac{1}{2}( \epsilon_{ij}(e_{kl}) E_i  E_j+ \mu_{ij}(e_{kl}) H_i H_j) ]  \nonumber \\
&=& -\partial_i S_i - \epsilon_{ij} E_i  \partial_t E_j -\mu_{ij}  H_i \partial_t H_j - \frac{1}{2}( \frac{\partial \epsilon_{ij}}{\partial e_{kl}} E_i  E_j+ \frac{\partial \mu_{ij}}{\partial e_{kl}} H_i H_j) \partial_t e_{kl}.
\end{eqnarray}
Evidently, the last term indicates the energy conversion between the elastic and electromagnetic fields.

Analogous to the analyzes in the above section, to guarantee the conservation of the total mechanic energy, then
\begin{eqnarray}
 &&\frac{d K }{dt} + \frac{d U^{el}}{dt} + \int_V [ \frac{1}{2}( \frac{\partial \epsilon_{ij}}{\partial e_{kl}} E_i  E_j+ \frac{\partial \mu_{ij}}{\partial e_{kl}} H_i H_j] \partial_t e_{kl} ) dv \nonumber \\
 &=& \int_V [ \partial_t(\frac{1}{2} \rho v_j v_j) + \frac{\partial u^{el}}{\partial e_{jk}} \partial_t e_{jk} +  \frac{1}{2}( \frac{\partial \epsilon_{ij}}{\partial e_{kl}} E_i  E_j+ \frac{\partial \mu_{ij}}{\partial e_{kl}} H_i H_j) \partial_t e_{jk} ] dv  \nonumber  \\
&=& \int_V [ \rho v_j \partial_t v_j + (\frac{\partial u^{el}}{\partial e_{jk}}  + \frac{\partial u^{em}}{\partial e_{jk}} ) \partial_t e_{jk} ] dv \nonumber  \\
&=& \int_V [ \rho \partial_t v_j - (\sigma^{el}_{jk}  + \frac{\partial u^{em}}{\partial e_{jk}}  )_{,k} ] v_j dv + \oint_A (\sigma^{el}_{jk}  + \frac{\partial u^{em}}{\partial e_{jk}} ) v_j n_k da  \nonumber \\
&=& 0.
\end{eqnarray}
Thus, the generalized stress is found to be
\begin{eqnarray}
\sigma^{ge}_{jk} = \sigma^{el}_{jk}  + \frac{\partial u^{em}}{\partial e_{jk}} = \frac{\partial (u^{el} + u^{em})}{\partial e_{jk}} . \label{eq:st2}
\end{eqnarray}
As a result, the constitutive equation (\ref{eq:st2}) guarantees the satisfaction of the law of conservation of energy. Furthermore, since externally applied stresses lead to variation of the  electromagnetic energy, thus the generalized Villari effects can be applied in sensors. In addition, from the above analysis, the electromagnetic energy can be treated as an extra contribution in the generalized elastic energy. In previous work,$^{\cite{Cahn:1985}}$ there were similar approaches of using the elastic energy as an extra contribution to the generalized chemical energy while studying diffusion under the influence of elastic fields. The generalized Villari effects in fact exemplify the conversion of energies owing to a cross-dependence in the energy terms, which is evidently different from the conversion of energies by nonpotential forces for the piezoelectric and piezomagnetic effects.  Moreover, the generalized Villari effects in fact arise from the dependence of the permittivity and permeability on the elastic stresses or strains. Since the Mauttecci effect refers to the arising of a helical anisotropy of the susceptibility of materials under the influence of a torque, then it should be classified as a specific example of the generalized Villari effects.

\section{Electrostriction and Magnetostriction}

It is known electrostriction and magnetostriction are induced by polarization and magnetization of the materials. Both of them are quadratic effects with constitutive equations being
\begin{eqnarray}
 e_{ij} &=& Q^e_{ijkl}P_kP_l \label{eq:ec}\\
e_{ij} &=& Q^m_{ijkl}M_k M_l \label{eq:mc}.
\end{eqnarray}
where $Q^e_{ijkl}$ and $Q^m_{ijkl}$ are electrostriction and magnetostriction coefficients, $P_k$ and $M_k$ are polarization and magnetization vectors.

Among the mechanisms of polarization in dielectric materials, it is believed the ionic and electronic polarization contribute mostly to electrostriction.$^{\cite{Sundar:1992}}$ That is, under the influence of an externally applied electric field, electrostriction mainly arises from the displacements of ions away from their equilibrium positions at the lattice and the distortion of  the electronic distribution around ions.  While for magnetostriction, boundary movements of the magnetic domains are believed to be the major cause.$^{\cite{Ekreem:2007}}$ Under the influence of an externally applied magnetic field $H_i$, domains with magnetic moments aligned with $H_i$ will grow. That is, magnetic moments around the domain boundary will rotate to the direction of $H_i$ and thus help the domain boundaries move forward. Such a process also simultaneously result in a dimensional change for the materials. During these processes, there are increases in energy and the energy increased in fact comes from the electric power source. According to the conservation of energy in electrodynamics as shown by Eqn (\ref{eq:emp}), the energy is in fact transferred into the bulk of the materials via the Poynting vector. Note that, the transfer of energy is always companied by the transfer of momentum simultaneously. Momentum is also needed during the processes of polarization and magnetization, e.g., when an ion is pushed out of its equilibrium position during the ionic polarization and when the magnetic moments start to rotate to help the domain grow. Furthermore, at the macroscopic level, materials under the influence of alternative fields usually vibrates, which indicates a gain of mechanic momentum. Then, where does this momentum (which is substantially the cause of electrostriction and magnetostriction) gained at both the microscopic and macroscopic levels come from? Evidently, it would be very beneficial if the law of conservation of momentum in electrodynamics is examined.

Here, electrostriction is mainly analyzed while magnetostriction is merely treated as an analogue. As we know, charged particles in electromagnetic fields are subject to the Lorentz force. The Lorentz force not only distorts the electronic distribution around ions, but also drives the ions away from their equilibrium positions during ionic polarization. Thus, it results in transfer of momentum to the ions. Then according to Newton's second law, the mechanic momentum $P^M_i$ obtained by all charged particles are
\begin{eqnarray}
\frac{dP^{M}_i}{dt} = \int_V (\rho E_i + \xi_{ijk} J^e_j B_k ) dv.
\end{eqnarray}
where the integrand at the right-hand side is the Lorentz force. In the following discussions, for simplicity, the permittivity and permeability are assumed constant, i.e. $\epsilon$ and $\mu$. Substitution of Eqns (\ref{eq:me1}) and (\ref{eq:me4}) into the above equation and further mathematical manipulations lead to the law of conservation of momentum in electrodynamics,$^{\cite{Jackson:1999}}$
\begin{eqnarray}
\frac{dP^{M}_i}{dt}+\frac{dP^{F}_i}{dt} = \oint_A T_{ij}n_j da, \label{eq:cm}
\end{eqnarray}
where the field momentum $P^{F}_i = \epsilon \mu \xi_{ijk} E_j H_k$ and $T_{ij}$, the Maxwell stress, is
\begin{eqnarray}
T_{ij} = \epsilon(E_i E_j-\frac{1}{2}E_n E_n \delta_{ij}) + \frac{1}{\mu} (B_i B_j - \frac{1}{2} B_n B_n \delta_{ij}). \label{eq:ms}
\end{eqnarray}
Evidently, as shown by Eqn (\ref{eq:cm}), both the mechanical momentum $P^M_i$ and the field momentum $P^F_i$ come from the momentum transferred by the Maxwell Stress. Note that $P^F_i$ is the momentum associated with the electromagnetic fields, while $P^M_i$ is the momentum obtained by ions and electrons during polarization and magnetization. As argued above, it is this mechanical momentum $P^M_i$ that causes dimensional changes, i.e., electrostriction and magnetostriction. Since this $P^M_i$ is transferred by the Maxwell stress $T_{ij}$, it is reasonable to argue that both electrostriction and magnetostriction are induced by $T_{ij}$. To be more specific, they are in fact induced by a portion of $T_{ij}$. First, assume the space considered is vacuum, then the mechanical momentum is zero since there are no particles; furthermore, the Maxwell stress acting on the surface, $T^0_{ij}$ (with $\epsilon$ and $\mu$ in Eqn (\ref{eq:ms}) replaced by $\epsilon_0$ and $\mu_0$ in the vacuum), then is totally responsible for the transfer of the field momentum $\epsilon_0 \mu_0 \xi_{ijk} E_j H_k$. Suppose now the space is occupied by materials. Then the variation in $P^F_i$ is $(\epsilon \mu -\epsilon_0 \mu_0) \xi_{ijk} E_j H_k$ which can be assumed minor for typical dielectric and paramagnetic substances. However, the mechanical momentum changes from 0 to $P^M_i$. Then the difference between $T_{ij}$ and $T^0_{ij}$ can be argued to be roughly the portion of stress being responsible of the transfer of $P^M_i$. Using $\epsilon=\epsilon_0(1+\chi_e)$; $\mu =\mu_0(1+\chi_m)$ and $P_i=\epsilon_0 \chi_e E_i$; $M_i=\mu_0 \chi_m H_i$, then this portion of stress is
\begin{eqnarray}
T^m_{ij} = T_{ij} -T^0_{ij} &=& \epsilon_0 \chi_e (E_i E_j-\frac{1}{2} E_n E_n \delta_{ij}) + \mu_0 \chi_m (H_i H_j - \frac{1}{2} H_n H_n \delta_{ij})  \label{eq:msem} \\
              &=& \frac{1}{\epsilon_0 \chi_e} (P_i P_j-\frac{1}{2} P_n P_n \delta_{ij}) + \frac{1}{\mu_0 \chi_m} (M_i M_j - \frac{1}{2} M_n M_n \delta_{ij}).
\end{eqnarray}
Since $T^m_{ij}$ induces both electrostriction and magnetostriction, then the corresponding strain is found to be
\begin{eqnarray}
e_{ij} = S_{ijkl} T^m_{kl} = \frac{S_{ijkl}}{\epsilon_0 \chi_e} (P_k P_l-\frac{1}{2}P_n P_n \delta_{kl}) +\frac{S_{ijkl}}{\mu_0 \chi_m} (M_k M_l - \frac{1}{2} M_n M_n \delta_{kl}). \label{eq:stn}
\end{eqnarray}
Here, strains are assumed to be infinitesimal strains. Compared with Eqns (\ref{eq:ec}) and (\ref{eq:mc}), the electrostriction and magnetostriction coefficients are found to be
\begin{eqnarray}
Q^e_{ijkl} = \frac{S_{ijkl}}{\epsilon_0 \chi_e}; \,\,\,\, Q^m_{ijkl} = \frac{S_{ijkl}}{\mu_0 \chi_m},
\end{eqnarray}
which qualitatively explains the resemblance of $Q^e_{ijkl}$ to $S_{ijkl}$ in a cubic crystal system.$^{\cite{Sundar:1992}}$

The Wiedemann effect found in 1858 is a specific example of magnetostriction. During the experiment, a ferromagnetic rod is placed within a solenoid. Then, a longitudinal magnetic field is induced by electric currents passing through the solenoid; and simultaneously, a circular magnetic field is induced by the electric currents passing through the rod. As a result, a torsion of the rod can be observed in this helical magnetic field. Previous analysis found that the angle of torsion is proportional to the longitudinal magnetic field and the magnitude of the current density passing through the rod, and inversely proportional to the shear modulus G.$^{\cite{Malyugin:1991}}$ Assume the rod is elastically isotropic. Then, using Eqn (\ref{eq:stn}), it is straight forward to show that the angle of torsion,
$e_{z\phi}=\frac{1}{2G\mu_0 \chi_m} M_z M_{\phi}$. Here, the longitudinal component of the magnetization, $M_z$, is proportional to the longitudinal magnetic field; the circumferential component, $M_{\phi}$, is proportional to the  the magnitude of the current density passing through the rod. Thus, this result is in agreement with the conclusion from the previous analysis. Furthermore, the rod is in fact subject to both electrostriction and magnetostriction. The induced longitudinal and circumferential strains are
\begin{eqnarray}
e_{zz} &=&  \frac{3(1+\nu)}{2E \epsilon_0 \chi_e} P_z P_z + \frac{3(1+\nu)}{2E \mu_0 \chi_m} (3 M_z M_z - M_{\phi} M_{\phi}), \\
e_{\phi \phi} &=&  \frac{3(1+\nu)}{2E \mu_0 \chi_m} (3 M_{\phi} M_{\phi} - M_{z} M_{z}).
\end{eqnarray}

In addition, from the analyzed above, the energy associated with electrostriction and magnetostriction can be assumed to be $\frac{1}{2}  T^m_{ij} e_{kl} = \frac{1}{2}  S_{ijkl} T^m_{ij} T^m_{kl}$. Here, the value of $T^m_{ij}$ is of the order of magnitude of the electromagnetic energy $u^{em}$ stored within the media. Since the strain resulted from electrostriction and magnetostriction is usually around $0.1\%$, then the kinetic energy obtained by mechanical vibrations is only roughly one thousands of $u^{em}$. Furthermore, according to Poynting's theorem, this energy is also transferred into the bulk of materials via the Poynting vector. Then when electrostriction and magnetostriction are considered, Poynting's theorem becomes
\begin{eqnarray}
J^e_i E_i &=&  -\partial_i (\xi_{ijk} E_j H_k ) - \partial_t [ \frac{1}{2}( \epsilon_{ij} E_i  E_j+ \mu_{ij} H_i H_j) + \frac{1}{2}T^m_{ij}S_{ijkl}T^m_{kl}].
\end{eqnarray}
For simplicity, Eqn (\ref{eq:msem}) is used for the substitution of $T^m_{kl}$. Mathematical manipulation, straightforward however tedious, found that the second and fourth equations of the Maxwell equations become
\begin{eqnarray}
 \xi_{1jk} \partial_j E_k &=& - \partial_t B_1 \nonumber \\
         &-&\frac{T^m_{kl} \mu_0 \chi_m}{\mu}[(S_{11kl}-S_{22kl}-S_{33kl})\partial_t B_1+2S_{12kl}\partial_tB_2,+2S_{13kl}\partial_tB_3]  \label{eq:emsb}\\
 J^e_1 &=& \xi_{1jk} \partial_j H_k + \partial_t D_1 \nonumber \\
&-& \frac{T^m_{kl}\epsilon_0 \chi_e}{\epsilon}[(S_{11kl}-S_{22kl}-S_{33kl})\partial_t D_1+2S_{12kl}\partial_tD_2,+2S_{13kl}\partial_tD_3] \label{eq:emsj}
 \end{eqnarray}
Permutation of indices $1\rightarrow2\rightarrow3$ leads to the rest two sets of equations. Detailed derivations can be found in the Appendix. It is evident these equations are highly nonlinear equations. Also note that, during the derivations above, the strains induced by electrostriction and magnetostriction are assumed to be infinitesimal strains. Thus Eqns (\ref{eq:emsb}) and (\ref{eq:emsj}) are only suitable for metallic or ceramic strictive materials. Nowadays, it is found that electrostriction and magnetostriction can induce definite and even large strains in polymer strictive materials. Then, the above analyzes must be combined with the theory of large deformations to tackle this problem.

Furthermore, note that the energy, $\frac{1}{2}  S_{ijkl} T^m_{ij} T^m_{kl}$, associated with electrostriction and magnetostriction, in fact depends on the electromagnetic fields only. Thus, it is a pure electromagnetic energy. However, it is this pure electromagnetic energy results in strains which are pure mechanical responses. In this case, energy associated with electromagnetic fields is directly converted into elastic energy, which is very different from the previously mentioned two cases (conversion of energies via the nonpotential forces or the cross-dependence of energy terms). When electromagnetic fields are applied, the charged particles within the substance are subject to the Lorentz force immediately. This leads to a distribution of the Maxwell Stress across the lattice of the substance and induces strains there, i.e., infinitesimal displacements of the charged particles away from their equilibrium positions. As a result, momentum is directly transferred by the Maxwell stress to the substance lattice. Simultaneously, the energy transferred from the power source by the Poynting vector is directly converted into elastic energy stored by the lattice. However, in this case, the energy associated with electrostriction and magnetostriction has no dependence on the elastic fields. Furthermore, since there is no inverse mechanisms, by which pure mechanic movements can result in a distribution of the Maxwell stress and induce the Lorentz force at the absence of electromagnetic fields, then there are no converse effects for electrostriction and magnetostriction discussed above. Note that, here electrostriction and magnetostriction only refer to the effects caused by the Maxwell stress, which do not include the generalized Villari effects.

In our previous work,$^{\cite{Zhou:2010}}$ it is assumed that the Maxwell stresses and strains can be linearly superposed with the elastic stresses and strains. The tensor product between the Maxwell stresses and the elastic strains, which include compositional strains as well, is the energy giving rise to electromigration. Here, it is argued the the tensor product, or a portion of the product, between the Maxwell stress and strain is the energy giving rise to electrostriction and magnetostriction.

In this section, both laws of conservation of momentum and energy are used to analyzes electrostriction and magnetostriction. It is argued that they are both induced by the Maxwell stress. The two dynamic equations of the Maxwell equations are found using Poynting's theorem; and the the energy associate with electrostriction and magnetostriction is found to be $\frac{1}{2}S_{ijkl} T^M_{ij}T^M_{kl}$.

\begin{figure}[htb]
\centering 
\epsfig{figure=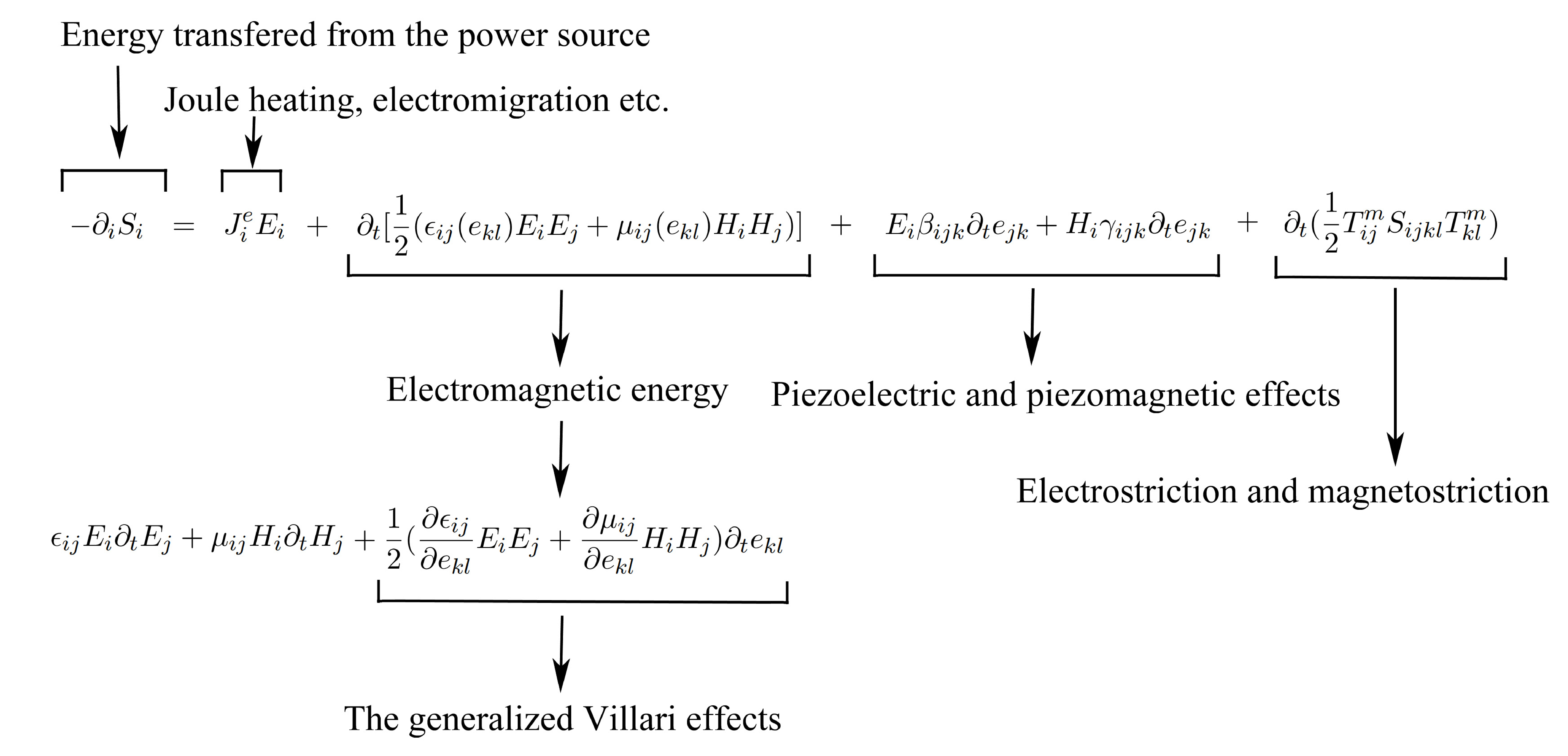,width=0.98\textwidth}
\caption{The conversion of energies for the coupling effects between the electromagnetic and elastic fields according to Poynting's Theorem.}
\label{fig:ecpt}
\end{figure}

\textbf{In the above three sections, the processes of conversion of energies are illustrated for the three types of coupling effects between electromagnetic fields and elastic fields. Using Poynting's Theorem, these conversion processes are summarized schematically  in Figure ~\ref{fig:ecpt} for the sake of clarity. In the following section, general coupling processes in electronic devices, which involves not only electromagnetic and elastic fields but also a compositional field, are analyzed}


\section{Formulation of General Coupling Processes in Electronic Devices}

Most physical processes usually involve several physical fields. The evolution of them leads to dissipation and conversion of energies simultaneously. For example, in certain electronic devices which use functionally gradient materials, electromagnetic, elastic and compositional fields evolve simultaneously. Thus, it is beneficial to analyze such general coupling processes here.

In our previous work, the diffusion process under the influence of elastic field is analyzed.$^{\cite{Zhou:2016}}$ The analysis there is extended to include the electromagnetic energy with permittivity and permeability being both dependent on the compositional field and strains, i.e., $u^{em} = \frac{1}{2} \epsilon_{ij} (c, e_{lm}) E_i E_j + \frac{1}{2} \mu_{ij}(c,e_{lm}) H_i H_j $. Consider a binary system with $c$ denotes the composition of one species. Then, in this system, the coupling processes cause the variation of following energies: the kinetic energy $k=\frac{1}{2}\rho v_l v_l$,  the elastic energy of $u^{el}(c,e_{lm})$, the chemical free energy of diffusion $f(c,e_{lm})$ and the electromagnetic energy $u^{em}$. In addition, a dissipation function $\frac{1}{m}J^c_i J^c_i$ is also needed for the diffusion process. Here, the law of conservation of energy is applied for each process respectively. For the mechanical movements, we have
\begin{eqnarray}
&&\frac{d}{dt} [\int_V(K+u^{el}+f+u^{em})dv ] + \int_V (E_i \beta_{ilm} \partial_t e_{lm} + H_i  \gamma_{ilm} \partial_t e_{lm}) dv \nonumber \\
       &=& \int_V [\frac{\partial}{\partial t}(\frac{1}{2} \rho v_l v_l)+ ( \frac{\partial u^{el}}{\partial e_{lm}} + \frac{\partial f}{\partial e_{lm}} + \frac{\partial u^{em}}{\partial e_{lm}} ) \partial_t e_{lm} ] dv + \int_V (E_i \beta_{ilm} \partial_t e_{lm} + H_i  \gamma_{ilm} \partial_t e_{lm}) dv  \nonumber \\
       &=& \int_V [ \rho v_l \partial_t v_l+ ( \frac{\partial u^{el}}{\partial e_{lm}} + \frac{\partial f}{\partial e_{lm}} + \frac{\partial u^{em}}{\partial e_{lm}}  + E_i \beta_{ilm} + H_i  \gamma_{ilm} )] v_{l,m} dv  \nonumber \\
       &=& \oint_A (\frac{\partial u^{el}}{\partial e_{lm}} + \frac{\partial f}{\partial e_{lm}} + \frac{\partial u^{em}}{\partial e_{lm}}  + E_i \beta_{ilm} + H_i  \gamma_{ilm} ) v_l n_m da  \nonumber \\
       &+& \int_V [\rho \partial_t v_l- ( \frac{\partial u^{el}}{\partial e_{lm}} + \frac{\partial f}{\partial e_{lm}} + \frac{\partial u^{em}}{\partial e_{lm}}  + E_i \beta_{ilm} + H_i  \gamma_{ilm} )_{,m} ] v_l dv = 0  \label{eq:el}
\end{eqnarray}
As a result, the equation of motion and the generalized stress tensor are found to be
\begin{eqnarray}
\rho  \partial_t v_l &=& \sigma_{lm,m} \nonumber \\
 \sigma^{ge}_{lm} &=& \frac{\partial u^{el}}{\partial e_{lm}} + \frac{\partial f}{\partial e_{lm}} + \frac{\partial u^{em}}{\partial e_{lm}}  + \beta_{ilm}E_i + \gamma_{ilm}H_i.
\end{eqnarray}
For the diffusion process, we have
\begin{eqnarray}
&&\frac{d}{dt} [\int_V(f+u^{el}+u^{em})dv ] + \int_V \frac{1}{m}J^c_i J^c_i dv \nonumber \\
&=& \int_V [ (\frac{\partial f}{\partial c} + \frac{\partial u^{el}}{\partial c} ++ \frac{\partial u^{em}}{\partial c}) \frac{\partial c}{ \partial t} dv + \int_V \frac{1}{m}J^c_i J^c_i dv \nonumber \\
&=&- \oint_A ( \frac{\partial f}{\partial c} + \frac{\partial u^{el}}{\partial c} + \frac{\partial u^{em}}{\partial c} ) J^c_i n_i da + \int_V [\frac{1}{m} J^c_i +(\frac{\partial f}{\partial c} + \frac{\partial u^{el}}{\partial c} +\frac{\partial u^{em}}{\partial c})_{,i} ] J^c_i dv=0. \label{eq:df}
\end{eqnarray}
Thus, the generalized flux equation is found to be
\begin{equation}
J^c_i = - m (\frac{\partial f}{\partial c} + \frac{\partial u^{el}}{\partial c} + \frac{\partial u^{em}}{\partial c})_{,i}.\\
\end{equation}
Then, substitution of the above equation into the law of conservation of mass leads to the governing equation of  the diffusion process. For the electromagnetic fields, we have
\begin{eqnarray}
 -\partial_i S_i & = & \partial_t u^{em} + J^e_i E_i + E_i \beta_{ijk} \partial_t e_{jk} + H_i  \gamma_{ijk} \partial_t e_{jk} \nonumber \\
 -\partial_i (\xi_{ijk} E_j H_k) & = & \partial_t [\frac{1}{2} \epsilon_{ij} (c, e_{lm}) E_i E_j + \frac{1}{2} \mu_{ij}(c,e_{lm}) H_i H_j] \nonumber\\
 &+& J^e_i E_i + E_i \beta_{ilm} \partial_t e_{lm} + H_i  \gamma_{ilm} \partial_t e_{lm} \nonumber \\
 -\partial_i (\xi_{ijk} E_j H_k) & = & \epsilon_{ij} (c, e_{lm}) E_i \partial_t E_j +  \frac{1}{2} \partial_t \epsilon_{ij} (c, e_{lm}) E_i E_j + \mu_{ij}(c,e_{lm}) H_i \partial_t H_j \nonumber \\
 &+& \frac{1}{2} \partial_t \mu_{ij}(c,e_{lm}) H_i H_j + J^e_i E_i + E_i \beta_{ilm} \partial_t e_{lm} + H_i  \gamma_{ilm} \partial_t e_{lm} \label{eq:ecce}\\
  \Longrightarrow 0&=& [\xi_{jik} \partial_j H_k + \epsilon_{ij} \partial_t E_j + \frac{1}{2} \partial_t \epsilon_{ij} E_j + J^e_i + \beta_{ilm} \partial_t e_{lm} ]E_i \nonumber \\
    &+& [\xi_{kji} \partial_k E_j + \mu_{ij} \partial_t H_j + \frac{1}{2}\partial_t \mu_{ij} H_j + \gamma_{ilm} \partial_t e_{lm}] H_i. \label{eq:ecmq}
\end{eqnarray}
As a result, the two dynamic equations of the Maxwell equations are found by setting the terms in the square brackets in the above equation equal to zero, and they are
\begin{eqnarray}
 && J^e_i = \xi_{ijk} \partial_j H_k - \partial_t D_i  + \frac{1}{2} (\frac{\partial \epsilon_{ij}}{\partial c} \partial_t c + \frac{\partial  \epsilon_{ij}}{\partial e_{lm}} \partial_t e_{lm}) E_j  \\
    && \xi_{ijk} \partial_j E_k = - \partial_t B_i + \frac{1}{2}(\frac{\partial \mu_{ij}}{\partial c} \partial_t c +\frac{\partial \mu_{ij}}{\partial e_{lm}} \partial_t e_{lm} ) H_j,
\end{eqnarray}
where $D_i =\epsilon_{ij}  E_j + \beta_{ilm} e_{lm}$ and $B_i= \mu_{ij} H_j +\gamma_{ilm} e_{lm} $, and extra terms are present taking into account the generalized Villari effects and the dependence of both $\epsilon$ and $\mu$ on the composition $c$. Furthermore, the following terms in Eqn (\ref{eq:ecce}) can be rewritten as
\begin{eqnarray}
&&\frac{1}{2} \partial_t \epsilon_{ij} (c, e_{lm}) E_i E_j + \frac{1}{2} \partial_t \mu_{ij}(c,e_{lm}) H_i H_j  \nonumber \\
&=& \frac{1}{2} ( \frac{\partial  \epsilon_{ij}}{\partial e_{lm}} \partial_t e_{lm}+\frac{ \partial \epsilon_{ij}}{\partial c} \partial_t c ) E_i E_j + \frac{1}{2} (\frac{\partial \mu_{ij}}{\partial e_{lm}} \partial_t e_{lm}+ \frac{\partial \mu_{ij}}{\partial c} \partial_t c ) H_i H_j \nonumber \\
&=& \frac{\partial u_{em}}{\partial e_{lm}} \partial_t e_{lm} + \frac{\partial u_{em}}{\partial c} \partial_t c.
\end{eqnarray}
Note that at the last step of the above equation, the two terms on the right-hand side are shown in Eqns (\ref{eq:el}) and (\ref{eq:df}) respectively. This guarantees that energy is conserved during the conversion of energies between the electromagnetic fields with the elastic and compositional fields, respectively. In previous work,$^{\cite{Zhou:2014}}$ it has already been shown energy is conserved during the conversion of energies between the elastic and compositional fields and thus it is omitted here.

In this section, general coupling processes, which are usually present in electronic devices and involve three physical fields, are discussed. With the energy formulation, governing equations for these fields are found. More importantly, it is shown these equations guarantee that energy is conserved during the process of energy conversions. This is no doubt an evident advantage of the energy formulation.

\section{Some Comments on Energy Conversions}

Most natural processes are in fact processes coupled among several physical fields. Thus, energy conversion is very common during these processes. According to the analyses above, energy can be converted via the following three different ways, i.e., work done by the nonpotential forces, cross dependence of energy terms and direct conversion as shown by electrostriction and magnetostriction. Compared with the first two ways, direct conversion is rare but it in fact reflects the substantial feature of energy conversion. In continuum media, energy is in fact stored via the electromagnetic interactions of ions and electrons. For examples, the elastic energy arises from variation of interactions among ions on lattice sites at the presence of strains, and so is chemical energy at the presence of compositional changes. As a result, when energy is converted directly, it is converted directly via the interactions of ions and electrons. In fact, for all three ways, energies are substantially converted via the interactions of ions and electrons. For examples, during the piezoelectric effect, externally applied pressures change the lattice distances and thus induce variation of interactions among ions at the lattice sites. And so is the generalized Villari effects.

In thermodynamics, the internal energy is a sum of all specific energies, such as the thermal energy, chemical energy, elastic energy and so on. These energies are all in fact stored via the interactions of ions and electrons within the substance. Thus, it is argued that there should be a general expression for the energy stored via the interactions of ions and electrons at the microscopic level. When this general expression is perturbed by the compositional field, then the variation is the chemical energy at the macroscopic level. When perturbed by the elastic or electromagnetic fields, the variations are the elastic or electromagnetic energies, respectively. It is believed determination of this general expression of the energy stored via the interactions of ions and electrons at the microscopic level can be quite beneficial for further understanding of energy conversions at the macroscopic level.

\section{Summary}

Mechanical movements, diffusion and electrodynamic process are all natural processes. For the simple, i.e., uncoupled processes, it is shown their governing or constitutive equations all satisfy the law of conservation of energy. Processes coupled among them substantially reflect the conversion of energy from one form into another. According to the law of conservation of energy, the loss of energy in one field must equal the gain of energy in another. This principle is used to analyze the reversible processes coupled between elastic and electromagnetic fields. It is found for the direct and converse piezoelectric and piezomagnetic effects which behave linearly, their constitutive equations guarantee that energy is conserved during the conversion of energies. Furthermore, it is argued that rather than associate with any specific energy terms, these processes are in fact associated with the work done by nonpotential forces. For the generalized Villari effects, using the law of conservation of energy, it is found while deriving the equation of motion for the elastic fields, the electromagnetic energy can be treated as an extra term in the generalize elastic energy. Thus, in this case, energy is converted via the cross dependence of the energy terms. While for electrostriction and magnetostriction, both the laws of conservation of momentum and energy are used to analyze them. It is found that they are induced by the Maxwell stress and their energy is purely electromagnetic. Since there are no inverse mechanism that pure mechanical movements can give rise to a distribution of the Maxwell stress across the lattice of substance, then it is argued that there are no converse effects for both of them.

General coupling processes in electronic devices which involves elastic, electromagnetic fields and diffusion are also analyzed. Both the equation of motion and the governing equation for diffusion are determined. Using Poynting's theorem, the two dynamic equations of the Maxwell equations are found with extra terms taking into account of the generalized Villari effects and the compositional dependence of permittivity and permeability. It is also shown these extra terms guarantee that energy is conserved during the energy conversion between the electromagnetic fields with the elastic and compositional fields.

Though the Lagrangian formulation is a traditional approach to determine the constitutive equations and construct governing equations for coupling processes. However, this formulation does not offer any physical insights into the conversion of energies during these processes. Furthermore, the phenomenological assumption of these energy terms, i.e. $\beta_{ijk} E_i e_{jk}$ and $\gamma_{ijk} H_i e_{jk}$, with the piezoelectric and piezomagnetic effects are also unreasonable. In this paper, the law of conservation of energy is used to construct governing equations directly for reversible processes coupled between electromagnetic fields and elastic fields. It is shown with this energy formulation, not only the conservation of energy is guaranteed for these coupling processes, but also the discussion of conversion of energies becomes straightforward. Furthermore, it also contribute to a better understanding of the underlying physical mechanisms of these processes. For an example, the conversion of energies during the piezoelectric and piezomagnetic effects are in fact achieved by the work of nonpotential forces. In a previous paper, PFVA with the free energy functionals has been used to study coupled irreversible processes and energy is also found to be conserved during the process of energy conversions.$^{\cite{Zhou:2014}}$ It is argued here this energy formulation when combined with PFVA can be extended to study most coupled natural reversible and irreversible processes so that their governing equations can strictly satisfy the law of conservation of energy and the second law of thermodynamics.

\section*{Acknowledgements}

We gratefully acknowledge the financial support from the National Natural Science Foundation of China (Grant No. 51201049).

\section*{Appendix}

For electrostriction and magnetostriction, Poynting's theorem becomes
\begin{eqnarray}
J^e_i E_i &=&  -\partial_i (\xi_{ijk} E_j H_k ) - \partial_t [ \frac{1}{2}( \epsilon_{ij} E_i  E_j+ \mu_{ij} H_i H_j) + \frac{1}{2}T^m_{ij}S_{ijkl}T^m_{kl}], \nonumber \\
&& T^m_{ij} = \epsilon_0 \chi_e (E_i E_j-\frac{1}{2} E_n E_n \delta_{ij}) + \mu_0 \chi_m (H_i H_j - \frac{1}{2} H_n H_n \delta_{ij}). \nonumber
\end{eqnarray}
Here, since $S_{ijkl}=S_{klij}$, then
\begin{eqnarray}
 &&\partial_t ( \frac{1}{2}T^m_{ij}S_{ijkl}T^m_{kl}) = S_{ijkl} T^m_{kl} \partial_t T^m_{ij}, \nonumber \\
 &&\partial_t T^m_{ij} = \epsilon_0 \chi_e (\partial_t E_i E_j+ E_i \partial_t E_j- E_n \partial_t E_n \delta_{ij}) + \mu_0 \chi_m (\partial_t H_i H_j + H_i \partial_t H_j - H_n \partial_t H_n \delta_{ij}). \nonumber
\end{eqnarray}
Expansion of $\partial_t T^m_{ij}$ leads to
\begin{eqnarray}
 \partial_t T^m_{11} &=& \epsilon_0 \chi_e (E_1 \partial_t E_1-E_2 \partial_t E_2-E_3 \partial_t E_3) + \mu_0 \chi_m (H_1 \partial_t H_1-H_2 \partial_t H_2-H_3 \partial_t H_3). \nonumber \\
 \partial_t T^m_{22} &=& \epsilon_0 \chi_e (- E_1 \partial_t E_1+E_2 \partial_t E_2-E_3 \partial_t E_3) + \mu_0 \chi_m (-H_1 \partial_t H_1+H_2 \partial_t H_2-H_3 \partial_t H_3). \nonumber \\
  \partial_t T^m_{33} &=& \epsilon_0 \chi_e (-E_1 \partial_t E_1-E_2 \partial_t E_2+E_3 \partial_t E_3) + \mu_0 \chi_m (-H_1 \partial_t H_1-H_2 \partial_t H_2+H_3 \partial_t H_3). \nonumber \\
  \partial_t T^m_{12} &=& \epsilon_0 \chi_e (E_2 \partial_t E_1+E_1 \partial_t E_2) + \mu_0 \chi_m (H_2 \partial_t H_1+H_1 \partial_t H_2). \nonumber \\
  \partial_t T^m_{13} &=& \epsilon_0 \chi_e (E_3 \partial_t E_1+E_1 \partial_t E_3) + \mu_0 \chi_m (H_3 \partial_t H_1+H_1 \partial_t H_3). \nonumber \\
  \partial_t T^m_{23} &=& \epsilon_0 \chi_e (E_3 \partial_t E_2+E_2 \partial_t E_3) + \mu_0 \chi_m (H_3 \partial_t H_2+H_2 \partial_t H_3). \nonumber
\end{eqnarray}
Thus,
\begin{eqnarray}
 &&\partial_t ( \frac{1}{2}T^m_{ij}S_{ijkl}T^m_{kl})= S_{ijkl} T^m_{kl} \partial_t T^m_{ij} \nonumber \\
 &&= S_{11kl}T^m_{kl}\partial_t T^m_{11} + 2 S_{12kl}T^m_{kl}\partial_t T^m_{12} + 2 S_{13kl}T^m_{kl}\partial_t T^m_{13} + S_{22kl}T^m_{kl}\partial_t T^m_{22} \nonumber \\
 && + 2 S_{23kl}T^m_{kl}\partial_t T^m_{23} + S_{33kl}T^m_{kl}\partial_t T^m_{33}, \nonumber \\
 &&= E_1 \epsilon_0 \chi_e (S_{11kl}T^m_{kl}\partial_t E_1 -S_{22kl}T^m_{kl}\partial_t E_1-S_{33kl}T^m_{kl}\partial_t E_1 + 2 S_{12kl}T^m_{kl}\partial_t E_2 + 2 S_{13kl}T^m_{kl}\partial_t E_3) \nonumber \\
 &&+ H_1 \epsilon_0 \chi_m (S_{11kl}T^m_{kl}\partial_t H_1 -S_{22kl}T^m_{kl}\partial_t H_1-S_{33kl}T^m_{kl}\partial_t H_1 + 2 S_{12kl}T^m_{kl}\partial_t H_2 + 2 S_{13kl}T^m_{kl}\partial_t H_3) \nonumber \\
 && + E_2 \epsilon_0 \chi_e (...) + H_2 \epsilon_0 \chi_m (...) + E_3 \epsilon_0 \chi_e (...) + H_3 \epsilon_0 \chi_m (...). \nonumber
\end{eqnarray}
Then the extra terms in Eqns (\ref{eq:emsb}) and (\ref{eq:emsj}) can be found and permutation of indices leads to the rest two sets of equations.

\end{document}